# Reciprocal Symmetric Boltzmann Function and Unified Boson-Fermion Statistics

Mushfiq Ahmad

Department of Physics, Rajshahi University, Rajshahi, Bangladesh

E-mail: mushfiqahmad@ru.ac.bd

Muhammad O. G. Talukder

Department of ETE and CSE, University of Development Alternative, Dhaka, Bangladesh

**Abstract**

The differential equation for Boltzmann's function is replaced by the corresponding discrete finite difference equation. The difference equation is, then, symmetrized so that the equation remains invariant when step d is replaced by –d. The solutions of this equation come in Boson-Fermion pairs. Reciprocal symmetric Boltzmann's function, thus, unifies both Bosonic and Fermionic distributions.

## 1. Introduction

We have seen that there exists a relation between reciprocal symmetry and spin[1]. It is, then, reasonable to expect that there exists a relation between reciprocal symmetry and quantum statistics. We shall study this below.

If the differential equation for the exponential function is replaced by the corresponding discrete symmetric difference equation, solutions come in pairs[2]. One of them corresponds to classical exponential function and the other is an oscillating function and has no classical analogue. We have seen that the other solution corresponds to Fermi-Dirac statistics[3]. We want to study in this paper how to exploit this pair solution to obtain a Boson-Fermion unified statistical relation.

## 2. Boltzmann's Finite Difference Equation

Classical Boltzmann's function $g$ satisfies the differential equation[4]

$$\frac{dg}{dH} = -wg = -w.\exp(-w.H) \tag{2.1}$$

The number $N^H$ of particles in energy level $H$ is given by Boltzmann's probability distribution relation[5]

$$N^H = N^0 \exp(-H.w) \tag{2.2}$$

Where $N^0$ is the number of particles when $H=0$

## 3. Planck's Radiation Formula

The formula for the distribution of light in a black body is given by[6]

$$I(\omega)dw = <H> \frac{\omega^2 d\omega}{\pi^2 c^2} \tag{3.1}$$

(2.2) together with Planck's hypothesis have been invoked to find $<H>$[7].

$$<H> = \frac{\eta\omega}{\exp(\eta\omega/kT) - 1} \tag{3.2}$$

(2.2) will lead to (3.2) if

$$H = n\eta\omega \tag{3.3a}$$

$$w = 1/kT \tag{3.3b}$$

## 4. Boltzmann's Symmetric Equation

To exploit its symmetry properties we replace differential equation (2.1) by the corresponding symmetric finite difference equation[8]

$$\frac{Dg_\pm^H}{D(H,\delta)} = -W \cdot g_\pm^H \tag{4.1}$$

where

$$\frac{Dg_\pm^H}{D(H,\delta)} = \frac{g_\pm^{H+\delta} - g_\pm^{H-\delta}}{2\delta} \tag{4.2}$$

By replacing (2.1) by (4.1) we have tacitly assumed $H = n.\delta$ where $n$ is an integer. The difference quotient (4.1) has the following symmetry under the change $\delta \to -\delta$

$$\frac{Dg_\pm^H}{D(H,-\delta)} = \frac{Dg_\pm^H}{D(H,\delta)} \tag{4.3}$$

We require that (4.1) should go over to (2.1) in the limit $\delta \to 0$

$$\frac{Dg_\pm^H}{D(H,\delta)} = -Wg_\pm^H \xrightarrow[\delta \to 0]{} \frac{dg}{dH} = -wg \tag{4.4}$$

With

$$g_+^H \xrightarrow[\delta \to 0]{} g \text{ and } W \xrightarrow[\delta \to 0]{} w \tag{4.4}$$

## 5. Boltzmann's Symmetric Function

(4.1) has solutions in pairs. One pair is

$$g_\pm^H = \exp\left(\frac{2\pi i}{\delta} s_\pm \, \mu \, w\right) H \tag{5.1}$$

Where $s_+$ is an integer and $s_-$ is a half-integer.

Both $g_+^H$ and $g_-^H$ satisfy (4.1) with

$$W = \frac{\sinh(w\delta)}{\delta} \tag{5.2}$$

$g_+^H$ and $g_-^H$ are related through reciprocity relation

$$g_+^H \cdot g_-^H = (-1)^{H/\delta} \tag{5.3}$$

## 6. Normalization

$g_+^H$ of (3.1) corresponds to classical Boltzmann's function. It is the classical function if we set $s_+ = 0$.

$g_-^H$ diverges for positive $H$ as $H \to \infty$. Therefore, normalization requirement[9] imposes the condition that $H$ must be negative.

## 7. Symmetric Number of Particles

If classical $g$ is replaced by symmetrized $g_+^H$ or $g_-^H$, the number of particles in energy level $H$ becomes

$$N_\pm^H = N^0 g_\pm^H \tag{7.1}$$

Where $N^0$ is the number of particles when $H=0$. If we set $N^0 = 1$ and $H = n.\delta$ (See comment following (4.2))

$$N_\pm^H = g_\pm^H = g_\pm^{n.\delta} \tag{7.2}$$

When $n=1$

$$N_\pm^\delta = g_\pm^\delta = \exp\left(\frac{2\pi i}{\delta} s_\pm \pm w\right)\delta \tag{7.3}$$

## 8. Symmetric Radiation Formula

We shall call $<H_\pm>$ the average energy corresponding to $<H>$ when the classical number of particles $N^H$ of (2.2) is replaced by the corresponding symmetric number $N_\pm^H$ as given by (4.2). Following the procedure[10] for finding (3.2), we find for $<H_\pm>$

$$<H_\pm> = \frac{\delta}{g_\pm^\delta - 1} \tag{8.1}$$

Using (5.1) and (3.3b), (8.1) becomes

$$<H_\pm> = \frac{\delta}{\pm\exp(\pm\delta/kT) - 1} = \frac{\pm\delta}{\exp(\pm\delta/kT) \mp 1} \tag{8.2}$$

## 9. Planck's Hypothesis

Planck's hypothesis has two parts: (a) Energy is discrete $H = n.\delta$ and that (b) $\delta = \eta\omega$. Part (a) has already been taken care of in sections 2 and 3. If we include part (b), (8.2) becomes

$$< H_{\pm} >= \frac{\pm \eta\omega}{\exp(\pm\eta\omega/kT) \mp 1} \tag{9.1}$$

## 10. Unified Boson-Fermion Statistics

The upper sign in (9.1) gives Planck's relation

$$< H_{+} >= \frac{\eta\omega}{\exp(\eta\omega/kT) - 1} \tag{10.1}$$

The lower sign in (9.1) gives the corresponding Fermi-Dirac relation[11]

$$< H_{-} >= \frac{-\eta\omega}{\exp(-\eta\omega/kT) + 1} = \frac{\eta\omega'}{\exp(\eta\omega'/kT) + 1} \tag{10.2}$$

Where $\omega' = -\omega$

## 11. Conclusion

Symmetric Boltzmann's function is enough to get both Bose-Einstein (10.1) and Fermi-Dirac (10.2) radiation relations from the corresponding classical relation[12] just as Planck's hypothesis is enough to obtain Planck's relation (3.2). To derive (10.2) we have not invoked Pauli Exclusion Principle.

---

[1] Mushfiq Ahmad. Reciprocal Symmetry and Origin of Spin. arXiv:math-ph/0702043v1
[2] Mushfiq Ahmad. Reciprocal Symmetry and Equivalence between Relativistic and Quantum Mechanical Concepts arXiv:math-ph/0611024v1
[3] Mushfiq Ahmad. Reciprocal Symmetric and Origin of Quantum Statistics. arXiv:physics/0703194v1
[4] Rudolf Kurth. Axiomatics of Classical Statistical Mechanics. Pergamon Press. 1960.
[5] http://www.answers.com/topic/boltzmann-distribution
[6] Feynman, Leighton, Sands. Feynman Lectures on Physics. Addison-Wesley Pub. Co.
[7] Feynman, Leighton, Sands. Feynman Lectures on Physics. Addison-Wesley Pub. Co
[8] Mushfiq AhmadReciprocal Symmetric and Origin of Quantum Statistics. http://www.arxiv.org/abs/physics/0703194
[9] Rudolf Kurth. Axiomatics of Classical Statistical Mechanics. Pergamon Press. 1960.
[10] Feynman, Leighton, Sands. Feynman Lectures on Physics. Addison-Wesley Pub. Co
[11] http://en.wikipedia.org/wiki/Fermi-Dirac_statistics

[12] Feynman, Leighton, Sands. Feynman Lectures on Physics. Addison-Wesley Pub. Co